\documentclass{article}

\usepackage{PRIMEarxiv}

\usepackage[utf8]{inputenc} 
\usepackage[T1]{fontenc}    
\usepackage{hyperref}       
\hypersetup{
    colorlinks=true,
    linkcolor=black,
    citecolor=black,
    urlcolor=blue,
    filecolor=black,
    pdftitle = {GenAI Workbench: AI-Assisted Analysis and Synthesis of Engineering Systems from Multimodal Engineering Data},
    pdfkeywords = {Model-Based Systems Engineering (MBSE); Generative AI; Digital Thread; System Architecture Analysis; Design Automation},
    pdfauthor = {H. Sinan Bank and Daniel R. Herber},
}
\usepackage{url}            
\usepackage{booktabs}       
\usepackage{amsfonts}       
\usepackage{amsmath}        
\usepackage{nicefrac}       
\usepackage{microtype}      
\usepackage{fancyhdr}       
\usepackage{graphicx}       
\graphicspath{{media/}}     

\usepackage{tikz}
\definecolor{gsblue}{HTML}{4285F4}
\DeclareRobustCommand{\gsicon}{%
	\begin{tikzpicture}[baseline=-0.35em]
	\draw[gsblue, fill=gsblue] (0,0) circle [radius=0.16];
	\node[white] at (0,0) {\fontfamily{phv}\selectfont\bfseries\tiny G};
	\end{tikzpicture}%
}

\newcommand{\scholarauthorA}{vU6oBhwAAAAJ}  
\newcommand{\scholarauthorB}{rn7gJxMAAAAJ}  
\newcommand{\scholarA}{\hspace{0.25em}\href{https://scholar.google.com/citations?user=\scholarauthorA}{\gsicon}}
\newcommand{\scholarB}{\hspace{0.25em}\href{https://scholar.google.com/citations?user=\scholarauthorB}{\gsicon}}

\title{GenAI Workbench: AI-Assisted Analysis and Synthesis of Engineering Systems from Multimodal Engineering Data
\thanks{\textit{\underline{Citation}}:
\textbf{H.S. Bank and D.R. Herber. GenAI Workbench: AI-Assisted Analysis and Synthesis of Engineering Systems from Multimodal Engineering Data. In Proceedings of the 2026 IISE Annual Conference, Arlington, TX, USA, May 16--19, 2026.}}
}

\author{
  H. Sinan Bank\scholarA\thanks{Corresponding author} \\
  Department of Systems Engineering \\
  Colorado State University \\
  Fort Collins, Colorado, USA \\
  \texttt{sinan.bank@colostate.edu} \\
  \And
  Daniel R. Herber\scholarB \\
  Department of Systems Engineering \\
  Colorado State University \\
  Fort Collins, Colorado, USA \\
  \texttt{daniel.herber@colostate.edu} \\
}

\begin{document}
\maketitle

\begin{abstract}
Modern engineering design platforms excel at discipline-specific tasks such as CAD, CAM, and CAE, but often lack native systems engineering frameworks. This creates a disconnect where system-level requirements and architectures are managed separately from detailed component design, hindering holistic development and increasing integration risks. To address this, we present the conceptual framework for the GenAI Workbench, a Model-Based Systems Engineering (MBSE) environment that integrates systems engineering principles into the designer's workflow. Built on an open-source PLM platform, it establishes a unified digital thread by linking semantic data from documents, physical B-rep geometry, and relational system graphs. The workbench facilitates an AI-assisted workflow where a designer can ingest source documents, from which the system automatically extracts requirements and uses vision-language models to generate an initial system architecture, such as a Design Structure Matrix (DSM). This paper presents the conceptual architecture, proposed methodology, and anticipated impact of this work-in-progress framework, which aims to foster a more integrated, data-driven, and informed engineering design methodology.
\end{abstract}


\section{Introduction} \label{s:intro}

Modern engineering design is hampered by the fragmentation of data and tools, where system-level requirements, functional architectures, and detailed physical designs often reside in disconnected environments \cite{hedberg2020using}. This separation creates information silos that break the digital thread, complicating verification and validation across the product lifecycle and increasing the risk of integration failures. The central motivation for this work is to bridge this gap by developing a cohesive and data-driven engineering methodology that moves toward the vision of ``MBSE 2.0''---an integrated, comprehensive, and intelligent paradigm for systems engineering \cite{zhang2025mbse20}.

Recent years have seen growing efforts to incorporate Artificial Intelligence (AI) into Model-Based Systems Engineering (MBSE) and engineering design tools. Researchers and practitioners recognize that current MBSE tools remain siloed and manual, motivating visions that deeply integrate AI across the system lifecycle. For example, Zhang et al. call for an AI-driven MBSE Co-Pilot to assist systems engineers, leveraging large language models (LLMs) for tasks such as model generation and analysis \cite{zhang2025mbse}. Initial prototypes have integrated GPT-4 into SysML modeling environments to auto-generate model elements from prompts \cite{longshore2024leveraging}, and other work has demonstrated that feeding an MBSE model into an LLM-based design assistant improves requirement generation and reduces hallucinations \cite{patel2024easing, garcia2024opm}.

Our central hypothesis is that by embedding AI-driven systems engineering capabilities directly into the designer's primary workflow, we aim to establish a unified digital thread that significantly enhances engineering effectiveness. To test this, we propose the GenAI Workbench, a \textit{conceptual framework} and integrated software environment that instantiates this next-generation engineering platform. It is critical to clarify at the outset that the GenAI Workbench is not presented as a commercial product, but rather as the methodological proof-of-concept required to validate the interoperability of multimodal data and generative synthesis workflows. This distinction is essential: the workbench serves as a scientific instrument to demonstrate that the proposed integration is computationally feasible and methodologically sound.

This paper details the conceptual architecture and proposed methodology of the GenAI Workbench, a system designed to enhance engineering effectiveness by embedding automated, system-level analysis and synthesis directly into the designer's daily workflow. The following sections review the state-of-the-art in AI for MBSE, describe the technical framework and AI-assisted workflow of the Workbench, discuss its anticipated impact on systems engineering practices with illustrative applications, and conclude with the broader implications and future work directions.

\section{Background and Related Work} \label{s:sota}

\textbf{Technical Approaches to Requirements Engineering with AI:} Automating requirements extraction has been a focal point for AI in systems engineering, with varying degrees of success. Transformer-based language models such as BERT and RoBERTa have been fine-tuned for requirement classification tasks, but they require labeled data and are limited in generating new content. The advent of GPT-3.5/GPT-4 and open models such as LLaMA-2 has enabled more flexible zero-shot or few-shot approaches. However, Li et al. found that GPT-based models achieved only about 45--52\% accuracy in extracting formal software specifications, often fabricating details or omitting information \cite{li2025specsllm}. Their work introduced a two-step ``annotation-then-conversion'' method, where one LLM first identifies candidate requirement sentences and another converts those into formal logic, significantly reducing errors. This decomposition helps mitigate the LLM's tendency to run off-track by focusing it on one sub-task at a time. While these accuracy rates are insufficient for production use without safeguards, they motivate the multi-stage refinement approach central to the proposed framework.

For vision-language models (VLMs) applied to engineering documents, Khan et al. present a hybrid framework that combines object detection (YOLO-based) with a transformer-based parser to read 2D mechanical drawings \cite{khan2025drawings}. By fine-tuning models such as Donut (a document image understanding model) on annotated drawings, they achieved 93.5\% F1-score in extracting structured information such as dimensions and tolerances. The TechMB dataset benchmarked multiple VLMs on technical drawing interpretation, finding that smaller open models struggled with complex drawings, whereas fine-tuned models can capture specific features \cite{garcia2023techmb}. This suggests that integrating a vision component to parse diagrams in PDF requirements is feasible, though specialized fine-tuned models currently provide more reliability than general-purpose multimodal LLMs such as GPT-4 Vision.

Traceability is another aspect where AI assists: commercial tools and emerging LLM-based Q\&A systems let engineers query large requirements documents in natural language, helping to trace dependencies or verify consistency. However, commercial MBSE platforms are only beginning to incorporate such AI features, making automated requirement extraction and linking particularly novel in practice.

\textbf{Graph-Based Approaches to System Architecture Integration:} In academia, researchers have explored using graph databases to unify multi-domain models. Schummer et al. describe storing SysML models in a Neo4j graph database to enable complex queries across systems, software, and simulation models \cite{schummer2022graphmbse}. Their approach facilitates analyses (e.g., finding paths, dependencies) that traditional SysML tools struggle with. They successfully stored a satellite SysML model in Neo4j, enabling queries to find which components satisfy certain requirements or to traverse functional chains. The advantages of graph databases include flexibility (easy to add new relationship types without altering a rigid schema) and performance on traversal queries. However, a standalone graph database introduces another technology to integrate and maintain, and graph query languages (e.g., Cypher) are less familiar to many engineers.

Intercax Syndeia provides graph-based linkages between SysML modeling tools (e.g., MagicDraw) and various engineering artifacts (e.g., CAD files, requirements in DOORS, Simulink models), essentially creating a federated graph of cross-domain relationships \cite{syndeia2021neo4j}. Another notable effort is the open-source project OpenMBEE (by NASA/JPL), which provides a model repository and view editor to keep documents and SysML models in sync---a step toward bridging human-readable docs and formal models \cite{openmbee2018}. These tools provide valuable deterministic linking at the artifact and parameter level (e.g., linking SysML blocks to CAD files, requirements to blocks). Unlike these, the proposed framework aims to link semantic requirements directly to the elements of physical viewpoint (e.g., geometric features, etc.) rather than CAD files as opaque artifacts, and uses AI-assisted synthesis from unstructured documents.

Hedberg et al. used a graph-based approach where every lifecycle artifact (requirements, CAD parts, process plans, etc.) is a node in a graph, and typed edges represent relationships such as ``refines'', ``implements'', or ``tests'' \cite{hedberg2020using}. This allowed them to trace data throughout the product lifecycle using graphs and maintain viewpoint interoperability. The GenAI Workbench applies similar principles: it uses unique identifiers to link artifacts across modalities, where the Design Structure Matrix (DSM) can be viewed as an adjacency matrix representation of a graph where components are nodes and interactions are edges, though the implementation uses a PLM's native database rather than a dedicated graph database.

\textbf{Multimodal Data Integration Challenges:}
Creating a unified digital thread that links semantic data (textual requirements), geometric data (CAD models), and relational data (system architecture graphs) is non-trivial. Data heterogeneity is a primary challenge: these data exist in very different forms---requirements in natural language text, CAD models in B-Rep or mesh formats, and system models as graphs or matrices. Best practice is to establish unique identifiers for key entities and use those as cross-references. Standards such as OSLC (Open Services for Lifecycle Collaboration) define a mechanism to link artifacts via URIs and REST APIs, ensuring traceability even across different tools \cite{oslc2021core}.

Semantic-geometry linking is particularly novel. One approach uses annotations in CAD: modern CAD formats (such as STEP or JT) allow metadata attributes on geometry. Some researchers refer to this as geometry-based requirements management, where requirements text is linked with specific geometry representing the required space or shape \cite{hamilton2018geometry}. The lesson from such approaches is that a clear information flow and tool integration is needed to manage these links, and that adding a physical viewpoint (e.g., geometry) into requirements verification.

\section{Methodology: A Conceptual Framework, GenAI Workbench} \label{s:methods}

The GenAI Workbench is designed as a conceptual framework that implements the principles of Model-Based Systems Engineering (MBSE) for multimodal data. It is not a single, monolithic application, but rather a methodology that comprises a set of interconnected tools. Together, these tools provide the capabilities for handling Document, Geometry, and Graph data. The toolchain has been architected to enhance modularity, reproducibility, and accessibility for the scientific community. Rather than a closed software product, it serves to operationalize the methodology, proving that the proposed workflows are potentially executable and not merely philosophical.

\subsection{System Architecture and Technical Foundations}

\textbf{PLM/CAD Backbone:} The GenAI Workbench requires a PLM/CAD platform that provides several critical capabilities: (1) a programmable API (e.g., Python) for geometry processing and automation, (2) support for standard CAD formats (e.g., STEP) for interoperability, (3) access to geometric data structures (B-rep) through an open geometry kernel for computational analysis, and (4) extensibility for custom workbenches and integrations. Commercial platforms such as Autodesk Fusion 360 and Siemens NX offer these capabilities, as do open-source alternatives. Through the platform's API and geometry kernel, the Workbench will parse STEP files, traverse B-rep structures, perform geometric queries, and conduct Boolean operations---providing the computational means to interpret and analyze the physical structure encoded within the Geometry component.

\textbf{Multimodal Data Model:} A system of persistent Unique Identifiers (UIDs) serves as the technical backbone of this integration architecture. Each distinct component is assigned a UID, which is explicitly associated with its respective semantic annotations (Document), geometric file (Geometry), and node representation (Graph). However, mere technical linkage via UIDs is insufficient to effectively bridge the multimodal semantic gap. To move beyond simple file association and enable true generative design, the integration logic incorporates Function and Behavior-based connections. These conceptual mappings allow the system to narrow down and interpret specific UIDs based on their semantic annotations (e.g., relating a specific `flange' geometry to its `attachment' function). To manage these associations efficiently and enable cross-modal queries, a centralized indexing mechanism is employed, maintaining the mapping between a component's UID and the location of its data within each modality's storage.

\begin{figure}[htbp]
    \centering
    \includegraphics[width=1\textwidth]{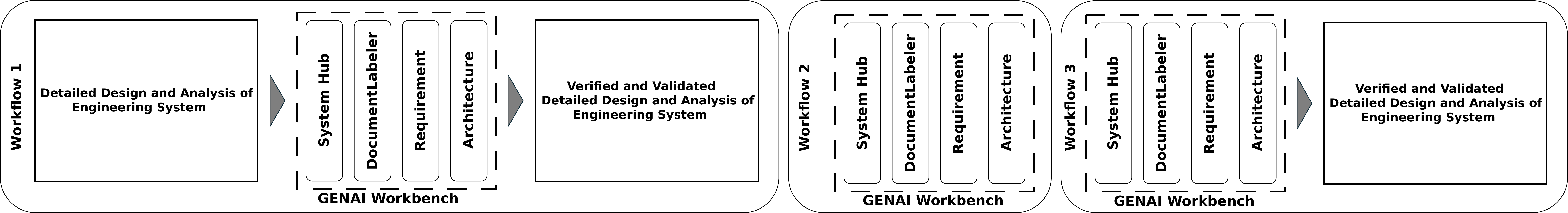}
    \caption{Proposed workflow methodology of the GenAI Workbench framework.}
    \label{fig:workflow}
\end{figure}

\textbf{Proposed AI-Assisted Workflow:} The GenAI Workbench introduces an AI-in-the-loop workflow that spans from initial requirements ingestion to system model generation to design visualization, all within the engineer's primary environment. The conceptual process proceeds through sequential stages as illustrated in Figure~\ref{fig:workflow}.

\textit{Document Ingestion:} The user provides source documents (e.g., a requirements specification in PDF/Word, design briefs, or standards). The Workbench's backend uses an LLM to parse these documents. First, it extracts raw text (and images if any)---using OCR for PDFs where needed. Then the LLM, guided by prompts, identifies requirement statements. Each requirement is saved as an object in the PLM (with fields such as text, source document reference, priority, etc.). Simultaneously, the system compiles a glossary of key terms to assist later steps. The output is a structured list of requirements in the MBSE Workbench, trace-linked to the source text so engineers can relate the provided text to the requirements.

\textit{Initial Architecture Synthesis:} With requirements extracted, the Workbench triggers the architecture generation. This involves interpreting the requirements collectively to propose an initial system breakdown. The LLM is prompted in the context of all requirements (possibly in summary form if too many). It identifies system components/subsystems by looking for nouns that represent physical or logical parts of the system. It also looks for relationships: e.g., if a requirement says ``The camera shall send data to the processing unit...'', we infer an interface between \textit{Camera} and \textit{Processor}. The LLM outputs a candidate list of components and pairwise interactions. We format this as a DSM: a matrix where rows and columns are the identified components and each cell indicates an interaction. This initial architecture is often a starting point---it may be incomplete or partially incorrect, but it provides a concrete model to discuss and refine.

\textit{Human-in-the-Loop Refinement:} The proposed requirements and architecture are presented to the user in the Workbench UI. The engineer can review each requirement (edit phrasing, correct misinterpretations, add missing ones)---these edits update the PLM objects. They can also review the DSM: perhaps the LLM suggested 5 subsystems, but the engineer knows of a sixth; they can add it via a UI, and likewise remove or modify connections. This is an important human-in-the-loop step to ensure the model captures domain knowledge correctly. Essentially, the AI jumpstarts the tedious work, and the human validates it.

\textit{Linking with CAD Design Environment:} Once the system architecture is confirmed at a high level, the designer proceeds with detailed design (e.g., creating or using existing 3D CAD models of each component). The GenAI Workbench, being integrated, ensures that for each component the user creates in CAD, there is a corresponding system element in the MBSE model. The key is establishing a one-to-one correspondence between system components in the architecture and CAD parts/assemblies in the design.

\subsection{Core Components}

\begin{figure}[t]
    \centering
    \includegraphics[width=1\textwidth]{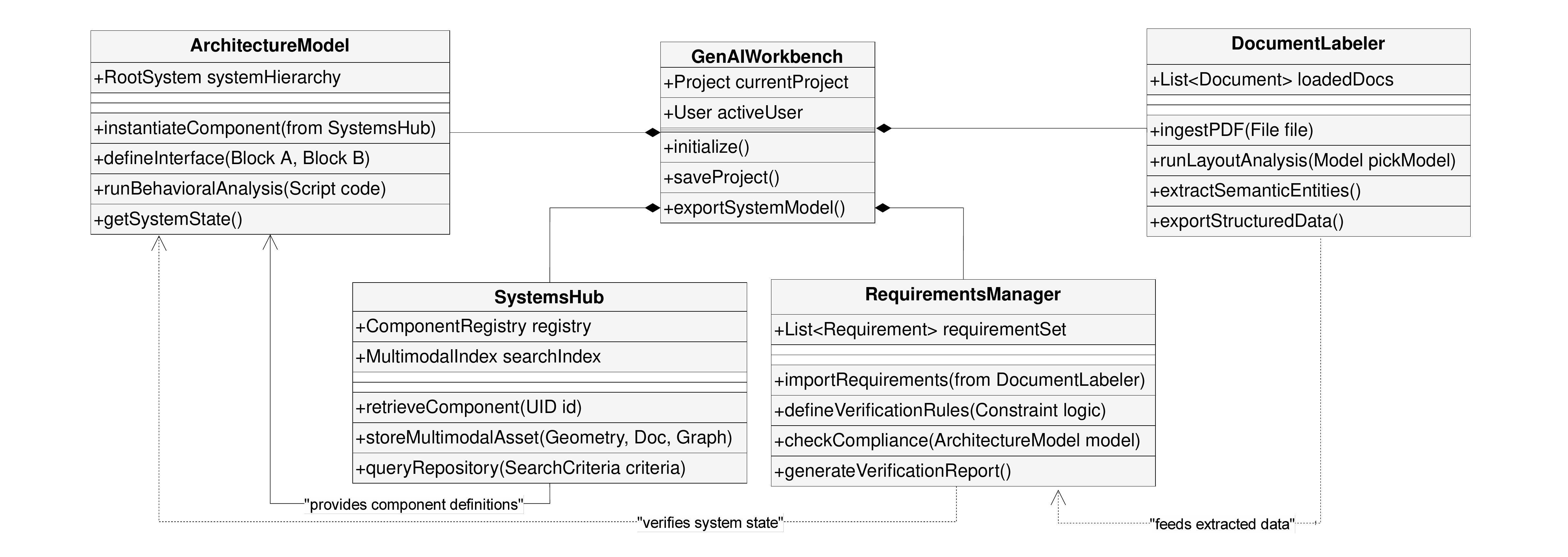}
    \caption{Backend architecture class diagram of the GenAI Workbench.}
    \label{fig:class_diagram}
\end{figure}

The workbench utilizes a set of core components architected to address the multimodal semantic gap and enable integrated MBSE environments. Figure~\ref{fig:class_diagram} shows the conceptual class diagram for the backend architecture. \textbf{Systems Hub (The Repository):} This component acts as the central repository for managing projects and the system hierarchy. It directly addresses the research question regarding the multimodal semantic gap by managing how diverse data modalities (text, geometry, graphs) are integrated. By ensuring that multimodal data is not siloed but semantically linked via unique identifiers, it provides the scalable data foundation required for the computational experiments \cite{bank2024catalogbank}. \textbf{DocumentLabeler (The Semantic Engine):} This component transforms raw documentation into structured semantic knowledge. It establishes the ``ground truth'' for training and validating the generative AI models. By converting static documents into machine-readable annotations, it enables the algorithmic exploration of extracting requirements and functions, effectively bridging the gap between natural language and engineering specifications \cite{bank2024catalogbank}. \textbf{Requirements (Verification):} Requirements are treated as first-class objects (with types, statements, and custom columns) that can be linked to specific parts of the architecture. This component connects the generative outputs (architectures) with engineering constraints, enabling the verification loop to mathematically verify if a synthesized system meets its functional goals. \textbf{Architecture (The Canvas):} This component provides the visual modeling environment to realize the System Architecture Generator. It supports hierarchical decomposition, allowing users to model systems, subsystems, and their interfaces via drag-and-drop. Furthermore, it integrates ``States and Actions'' to model behavior (e.g., state machines, Python analysis scripts).

\subsection{Verification and Validation within the Framework}

The workbench's methodology is designed to support continuous V\&V through a hierarchical verification system. The verification process operates in two proposed phases. \textbf{Phase 1: Cross-Modal Compatibility Checks.} The rule engine evaluates compatibility checks across modalities to ensure consistency between different data representations. \textit{Geometric compatibility} checks examine spatial relationships between connected parts, verifying alignment and dimensional consistency. \textit{Functional/Behavioral compatibility} checks compare relevant parameters, capacities, or operating ranges extracted from document annotations. \textit{Relational compatibility} checks leverage the graph context to ensure connection types are valid based on component functions and system-level constraints. \textbf{Phase 2: Formal Verification via Symbolic Methods.} To extend verification beyond pairwise checks and address the validity of system behaviors and sequences, we propose incorporating formal methods approaches. Natural language engineering constraints from the Document component could be translated into formal specifications using symbolic AI techniques. The verification process would then evaluate system models against these derived specifications to ensure behavioral correctness and temporal properties.

\section{Anticipated Results and Potential Applications} \label{s:impact}

\textbf{Realizing the Digital Thread:} The workbench's architecture is designed to make the ``digital thread'' a practical reality rather than a theoretical concept. Through the UID-based integration infrastructure, every requirement would be explicitly linked to the system components it constrains, every component linked to its geometric representation, and every interface traceable to its source documentation. This automated linking aims to ensure end-to-end traceability: an engineer could click on a CAD feature and immediately see the requirements it satisfies, or query ``show me all components affected by this requirement change'' and receive an accurate, automatically-maintained answer.

\textbf{Accelerating Early-Stage System Architecture Exploration:} The AI synthesis capability represents a potential paradigm shift for system architects. Instead of spending weeks manually decomposing requirements into components and interfaces, architects could rapidly generate and iterate on multiple candidate architectures within hours. This acceleration is critical in the early phases of a project, where decisions have the highest leverage: exploring a broader design space early can identify superior architectures that would otherwise be missed due to time constraints. The workbench enables this by treating architecture generation as a computational task that can be parameterized, repeated, and compared.

\begin{figure}[b]
    \centering
    \includegraphics[width=1\textwidth]{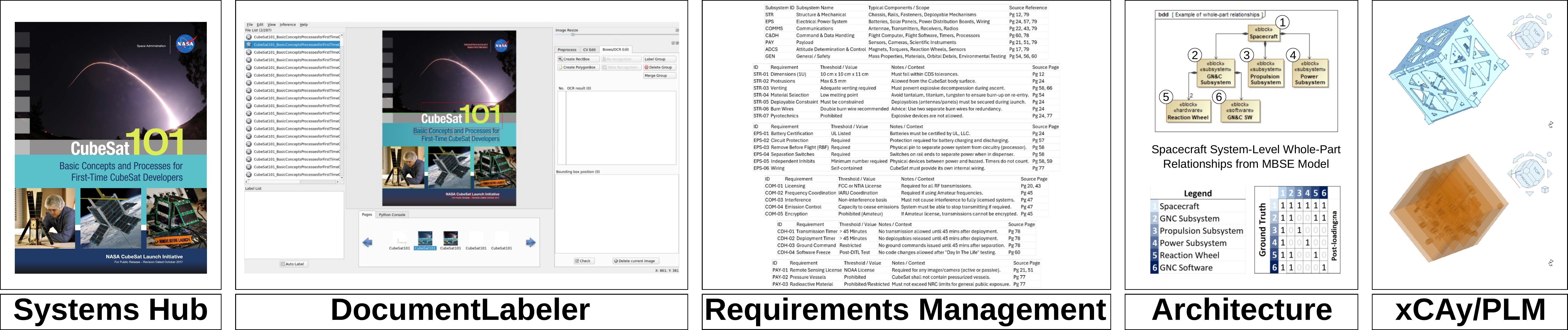}
    \caption{Illustrative example of the GenAI Workbench applied to a CubeSat design scenario.}
    \label{fig:cubesat}
\end{figure}

\subsection{Illustrative Domain Applications}

The GenAI Workbench methodology could be applied to complex systems across multiple domains:

\textbf{Aerospace (CubeSat Design):} NASA's CubeSat missions provide a great case, as illustrated in Figure~\ref{fig:cubesat}---typically small teams, lots of requirements (from launch vehicle constraints to payload functions), and tight mass/power budgets \cite{nasa2017cubesat101}. If GenAI Workbench were applied, the process of building the system model would be quicker and more thorough (especially in capturing all interface requirements from launch provider documents). We anticipate benefits such as ensuring no subsystem (comms, ADCS, power) violates system power or mass constraints by continuous checks.

\section{Conclusion and Future Work}\label{s:conclusion}

This paper has presented the conceptual framework for the GenAI Workbench, a proposed integrated MBSE environment that addresses the fragmentation between system-level and component-level engineering. The framework's key contributions include: (1) a unified multimodal data model based on UIDs that links semantic, geometric, and relational data; (2) three proposed AI-assisted workflows for analysis, synthesis, and generation of system models; (3) a hierarchical V\&V framework combining cross-modal rules with formal verification; and (4) an implementation strategy within an open-source stack centered on available geometry kernels.

Future work focuses on three directions: (1) \textbf{Implementation and Validation:} Completing the implementation of the proposed workflows and conducting formal user studies and case studies to empirically quantify the anticipated benefits; (2) \textbf{Deepening the Physical-Semantic Bridge:} Extending the geometry processing toolkit to utilize learned embeddings by fine-tuning Vision Language Models (VLMs) on engineering datasets; and (3) \textbf{Agentic Synthesis and Verification:} Integrating agentic workflows where the AI can autonomously execute Python-based analysis scripts within the workbench to verify compatibility, moving beyond the current human-in-the-loop paradigm toward more autonomous verification.

\bibliographystyle{unsrt}
\bibliography{references}

\end{document}